\documentclass[modern]{aastex701ed}
\NewPageAfterKeywords
\begin{document}

\title{A New Determination of the Mass of NGC 3603-A1: \\ the Most Massive Binary Known?}

\author[0000-0001-6563-7828]{Philip Massey}
\affiliation{Lowell Observatory, 1400 W Mars Hill Road, Flagstaff, AZ, 86001, USA}
\affiliation{Department of Astronomy and Planetary Science, Northern Arizona University, Flagstaff, AZ, 86011-6010, USA}
\email[show]{phil.massey@lowell.edu}

\author[]{Sarah Bodansky}
\altaffiliation{NSF REU student 2020}
\affiliation{Lowell Observatory, 1400 W Mars Hill Road, Flagstaff, AZ, 86001, USA}
\affiliation{Department of Astronomy and Planetary Science, Northern Arizona University, Flagstaff, AZ, 86011-6010, USA}
\affiliation{Department of Astronomy, University of Massachusetts, Amherst, MA, 01003, USA}
\email{sbodansky@umass.edu}

\author{Laura R. Penny}
\affiliation{Department of Physics and Astronomy, College of Charleston, 208 J.C. Long Building, 9 Liberty Street, Charleston, SC 29424, USA}
\email{PennyL@cofc.edu}

\author[0000-0003-2535-3091]{Nidia I. Morrell}
\affiliation{Las Campanas Observatory, Carnegie Observatories, Casilla 601, La Serena, Chile}
\email{nmorrell@carnegiescience.edu}

\author[0000-0002-5787-138X]{Kathryn F. Neugent}
\affiliation{Lowell Observatory, 1400 W Mars Hill Road, Flagstaff, AZ, 86001, USA}
\email{kathrynneugent@gmail.com}

\begin{abstract}

The star NGC~3603-A1 has long been known to be a very massive binary, consisting of a pair of O2-3If*/WN5-6 stars, which show Wolf-Rayet-like emission due to their luminosities being near the Eddington limit.   The system has been poorly characterized until now, due to the difficulties of obtaining reliable radial velocities from broad, blended emission lines and
the extreme crowding in the cluster.  However, previously unpublished archival HST/STIS spectra revealed
that some of the upper Balmer lines (seen in absorption) are well separated at favorable orbital phases, prompting us to obtain our own  carefully-timed new HST/STIS spectra, which we have analyzed along with the older data. Radial velocities measured from these spectra allow us to obtain an orbit for this 3.77298 day system.   We also used archival STIS imaging of the cluster to obtain a more accurate light curve for this eclipsing system, which we then modeled, yielding the orbital inclination and providing values for the stellar radii and temperatures. Together, these data show that the NGC~3603-A1 system consists of a $93.3\pm11.0 M_\odot$ O3If*/WN6 primary with
an effective temperature of 37,000~K, and a $70.4\pm9.3M_\odot$ O3If*/WN5 secondary that is slightly
hotter, 42,000~K.  Although a more massive binary is known in the LMC, NGC~3603-A1 is as massive as any binary known in our own Galaxy for which a direct measurement of its mass has been made. The secondary has been spun up by mass accretion from the primary, and we discuss the evolutionary status of this intriguing system.

\end{abstract}

\section{Introduction}

Understanding the evolution of massive stars requires understanding binary massive stars, as most massive stars are found in binaries.  At least 40\% of O stars are found in {\it close} binary systems  \citep{garmany,sana13}, and the percentage increases to 70\% or more if longer period systems are included (\citealt{2009AJ....137.3358M}; see also \citealt{2022A&A...667A..44G}). \citet{2012ASPC..465..275G} has argued that if undetected binaries are accounted for, the percentage of O-type stars in binaries might be 100\%, and \citet{sana12} has suggested that at least 70\% of O stars may undergo interactions during their lifetimes.   But even if that percentage is as low as 40-50\%, that is a significant part of the population.  The end result of massive star evolution is a neutron star or black hole, and it is the merger of binary pairs of such objects that lead to gravitational waves (see, e.g., \citealt{2013ApJ...779...72D,2017ApJ...846..170T}). Thus, understanding the properties and evolution of massive binaries is crucial to this new window on the Universe.

Studying massive stars in binaries is important for another, even more fundamental reason: it is the only direct way of getting a star's mass.  If radial velocities can be measured for both components, and if an orbital inclination can be determined (either from eclipses or from an interferometric ``visual" orbit),  then the masses can be determined simply from Kepler's laws.
In order to find the most massive binaries, one needs to look at clusters still young enough ($<$2-3~Myr) that their massive stars have not yet evolved and died, and rich enough to contain these rare objects.  NGC~3603 is the closest giant H {\sc ii} region, and contains one of the largest collections of massive stars known in the Milky Way, with $\sim$50 known O-type stars, including a dozen of the hottest and most luminous stars known \citep{1995AJ....110.2235D, 2008AJ....135..878M}.  Within the Local Group, only the R136 cluster possibly outstrips NGC 3603 in its massive star content (e.g., \citealt{1994ApJ...436..183M,MH98, 2014A&A...564A..40W}).

\begin{figure}
\epsscale{0.8}
\plotone{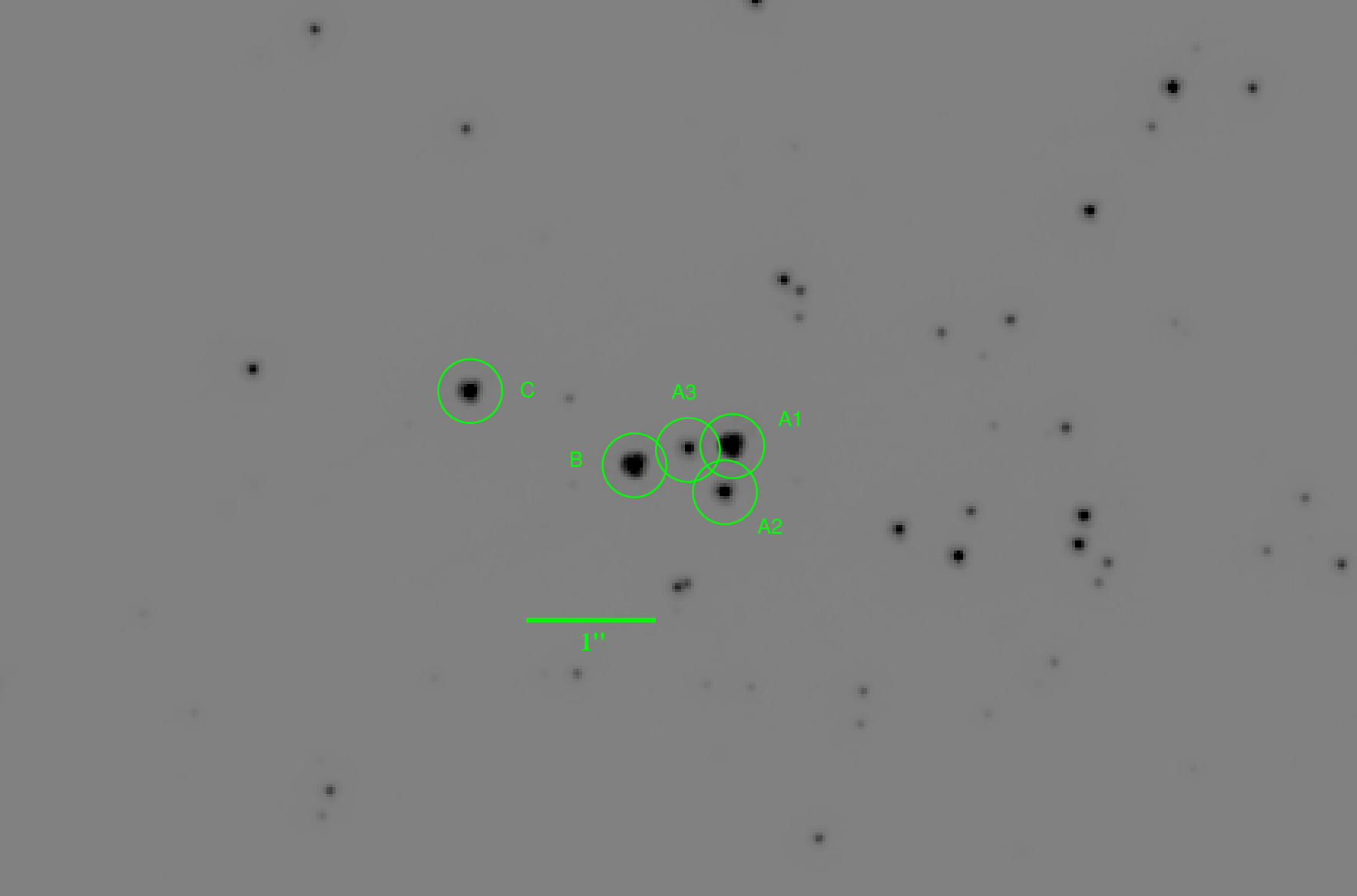}
\caption{\label{fig:central} The region around NGC 3603-A1.   The circles have diameters of 0\farcs25; the neighboring components A2 and A3 have separations from A1 of about one-third of an arcsecond. The image was obtained with the High Resolution Camera of the Advanced Camera for Surveys (ACS) on HST; it is a small part of an 8 s drizzled exposure obtained through the F550W filter as part of Program ID 10602 (PI: Maiz Apellaniz).  Components A1, B, and C were all called ``WN6+abs" by \citet{1995AJ....110.2235D}, although we would call them O2-3I*/WN5-6 ``slash" stars today; component A2 is classified as an O3~V, and A3 as an O3~III(f*) \citep{1995AJ....110.2235D}. North is up and east is to the left.
}
\end{figure}

NGC 3603-A1 is the brightest component of
the ``fuzzy" star HD 97950, located at the center of the cluster.   HD 97950 was described as a ``trapezium-like" system by \citet{1973ApJ...182L..21W}, who partially resolved it photographically in good seeing.    \citet{1984ApJ...284..631M} argued that the central object consisted of several Wolf-Rayet (WR) and O-type stars, finding that
the dominant WR spectrum showed radial velocity variations with a 3.7720 d period.  The system was resolved by speckle
observations  \citep{1986AA...167L..15H}, with components A1, A2, A3, and B cleanly separated. \citet{2004AJ....128.2854M} obtained Hubble Space Telescope (HST) J-band photometry with the Near Infrared Camera and Multi-Object Spectrometer (NICMOS) which showed that A1 eclipsed with the same ephemeris as  the radial velocity variations, suggesting that actual masses could be found for this extremely luminous (and hence presumably extremely massive) object.  The observational difficulty is, of course, the extreme crowding.  A small portion of a HST/ACS image is shown in Figure~\ref{fig:central}.

Using adaptive optics (AO) on the VLT, \citet{2008MNRAS.389L..38S} were able to derive a preliminary orbit from near-IR (NIR) spectroscopy, suggesting that NGC 3603-A1 is the most massive binary known, with masses of 116$\pm$31 $M_\odot$ and 89$\pm$16 $M_\odot$.
The large errors on the masses are due to the radial velocity curve of the secondary being poorly defined.  They  found that the NIR spectrum of NGC 3603-A1 consists of broad, strong emission lines of H and  He\,{\sc ii}.  These features are similar to those found in {\it bona-fide} WR stars, but are also found in the most luminous {\it unevolved} massive stars as first recognized by \citet{deKoterR136} and \citet{MH98}. These stars are now classified as O2-3If*/WN5-6, and their strong, broad emission lines are formed in optically thick stellar winds due to their luminosities being so close to their Eddington limits (e.g., \citealt{1998MNRAS.296..622C,2011A&A...535A..56G,1998MNRAS.296..622C,2019A&A...627A.151S,2020MNRAS.499..873S}). These ``Of-type stars on steroids" can be distinguished from ``real" (evolved) WRs  by the presence of hydrogen and their much higher visual luminosities (e.g., \citealt{MH98}).  For NGC~3603-A1, the width and blending of these broad lines resulted in large residuals  in the fit of the secondary's radial velocity curve.

In this paper we analyze radial velocity measurements and improved photometry using new and archival HST data in order to determine more accurate physical parameters for this interesting system. In Section~\ref{Sec-obs} we
describe the observations and measurements, and in Section~\ref{Sec-analysis} we present the new orbit solution and properties.
We discuss the significance of our findings in Section~\ref{Sec-discussion}.

\section{Observations and Reductions}
\label{Sec-obs}

Both the photometry and spectroscopy used here were  obtained using the Space Telescope Imaging Spectrograph (STIS)
on HST, and were obtained after the 2009 Servicing Mission 4 which resurrected the instrument after a 2-year hiatus due to a power supply failure.    The instrument is described by \citet{1998PASP..110.1183W}, and
information can be gleamed from many on-line web pages\footnote{See, e.g., \href{https://www.stsci.edu/hst/instrumentation/stis/instrument-design/detectors/}{https://www.stsci.edu/hst/instrumentation/stis/instrument-design/detectors/} and \href{https://www.stsci.edu/hst/instrumentation/stis}{https://www.stsci.edu/hst/instrumentation/stis}.} as well as the current version of the instrument manual \citep{2024stis.rept....5R}. The STIS CCD is a thinned, backside-illuminated SITe 1024$\times$1024 detector with 21$\mu$m pixels and a read-noise of 6.2 e$^{-1}$. The image scale is 0\farcs05078 pixel$^{-1}$, resulting in a field of view of 52\arcsec $\times$ 52\arcsec. 

Originally installed in 1997, time spent in the space environment has not been kind to the instrument.  Damage from trapped particle radiation in the Van Allen belts and cosmic rays has led to significant loss of charge transfer efficiency (CTE), particularly when the background is low \citep{2022acs..rept....5B}.  Radiation damage has also resulted in multiple ``hot pixels,"  pixels with very
high dark current \citep{1997hstc.work..120H}. These are temporarily repaired by occasionally warming the instrument.  In addition, cosmic ray events at the 4$\sigma$ level occur at a rate of about 25-30 pixels s$^{-1}$ according to \citet{1997hstc.work..120H}.
These challenges are usually met by a combination of sophisticated observing strategies and reduction techniques, as described
below.

\subsection{Imaging, Photometry, and the Orbital Ephemeris}
\label{Sec-photometry}

Photometry allows us to revise the orbital ephemeris, necessary for planning our own HST spectroscopy
(Section~\ref{Sec-spectra}) and is the means by which we can calculate the orbital inclination (Section~\ref{Sec-laura}).

The photometry was all done with 0.1~s unfiltered STIS exposures obtained
between 2010 March 22 to 2010 May 16 (Program ID 11626, PI: Massey) 
and between 2012 June 16
through 2012 August 21 (Program ID 12615, PI: Schnurr).
The first of these programs was aimed at determining the upper end of the
initial mass-function of NGC~3603, and consisted of 28 orbits during Cycle 17,
distributed over 8 visits of 4-5 orbits each.  Four dithered 0.1~s  images were obtained at the start and end of
each visit using the standard STIS-CCD-BOX pattern. 
The second program was aimed at obtaining phase-resolved spectroscopy of NGC~3603-A1 (discussed below), and consisted
of 14 orbits during Cycle 19, with each orbit done as a separate visit.
Four similarly dithered  0.1~s images were taken at the start of each of the 14 visits. 

The images were all taken through the ``50CCD" aperture, a clear, unvignetted opening.  Since
there is no filter, the wavelength coverage is set by the detector sensitivity, which extends from 
2000~\AA\ to 10,300~\AA, peaking from about 4000~\AA\ to 7500~\AA\ (see Figure 5.1 in \citealt{2024stis.rept....5R}).

A major challenge was applying correction for CTE losses, given the 0.1 s exposure times resulting in essentially
no background. We used the pixel-based {\sc stis\_cti} software package, which was based upon the empirical 
scheme developed by \citet{2010PASP..122.1035A} for HST/ACS data.

We measured the photometry on the individual exposures; i.e., no effort was
made to combine the four dithered exposures into one.  If photometry from one of the four exposures was discrepant 
(attributable either to a hot pixel or a cosmic ray) we removed it from the analysis. Photometry was done through a 3-pixel radius
aperture after correcting for CTE losses.   Sky values were found by taking the mode of the values
in an annulus that extended from a radius of 5 to 10 pixels.  The final magnitude is in the STMag
system, and was determined by multiplying the counts~s$^{-1}$ by the value of {\sc photflam} in the image header in order to obtain the flux, f$_\lambda$, (in units of erg~cm$^{-2}$~s$^{-1}$~\AA$^{-1}$), and then converting to STMag = $-2.5 \log f_\lambda - 21.1$.  The individual {\sc photflam} values varied by
0.01~mag, indicating a slight decrease in sensitivity over the two and a half years covered by the observations.  The final photometry is listed in Table~\ref{tab:phot}.
\begin{deluxetable}{l c c c}
\tablecaption{\label{tab:phot} Photometry of NGC~3603-A1}
\tablehead{
\colhead{HJD}
&\colhead{Phase\tablenotemark{a}} 
&\colhead{STMag}
&\colhead{$\sigma_{\rm STMag}$}
} 
\startdata
2455278.3210 & 0.956 &  11.832 & 0.006 \\
2455278.3218 & 0.956 &  11.838 & 0.005 \\
2455278.3226 & 0.956 &  11.842 & 0.005 \\
2455278.6401 & 0.040 &  11.837 & 0.005 \\
2455278.6409 & 0.041 &  11.850 & 0.006 \\
2455278.6418 & 0.041 &  11.843 & 0.004 \\
2455278.6426 & 0.041 &  11.840 & 0.005 \\
2455332.3900 & 0.286 &  11.503 & 0.005 \\
2455332.3908 & 0.287 &  11.502 & 0.005 \\
2455332.3916 & 0.287 &  11.483 & 0.006 \\
\enddata
\tablecomments{Table 1 is published in its entirety in the machine-readable format.
      A portion is shown here for guidance regarding its form and content.}
\tablenotemark{a}{Computed using  P=3.77298 days and T0=2456161.3650.}
\end{deluxetable}

We then used these data to determine the orbital ephemeris.  \citet{1984ApJ...284..631M}  derived a period of 3.7720$\pm$0.0003 d 
and a zero-point T0 of 2444258.39$\pm$0.10 based on radial velocity measurements. (Zero phase was chosen to correspond with the
most massive star in inferior conjunction.)  This zero-point and period was later adopted by  \citet{1985ApJ...295..109M}
in phasing the velocities of the emission lines of the unresolved central objects.  \citet{2004AJ....128.2854M} used their NICMOS J-band photometry of A1 to revise the period slightly to 3.7724 d and update the zero-point to T0=2450513.520.  The same convention for the zero-point was maintained, possibly for convenience given that the two eclipses were of similar depth. \citet{2008MNRAS.389L..38S}  adopted the 3.7724 d period
from \citet{2004AJ....128.2854M}, and revised the zero-point to T0=2453765.75.  We note that all of these phase zero-points are consistent to integral cycles; i.e., they are based on the more massive star being at inferior conjunction. 

\begin{figure}
\epsscale{0.8}
\plotone{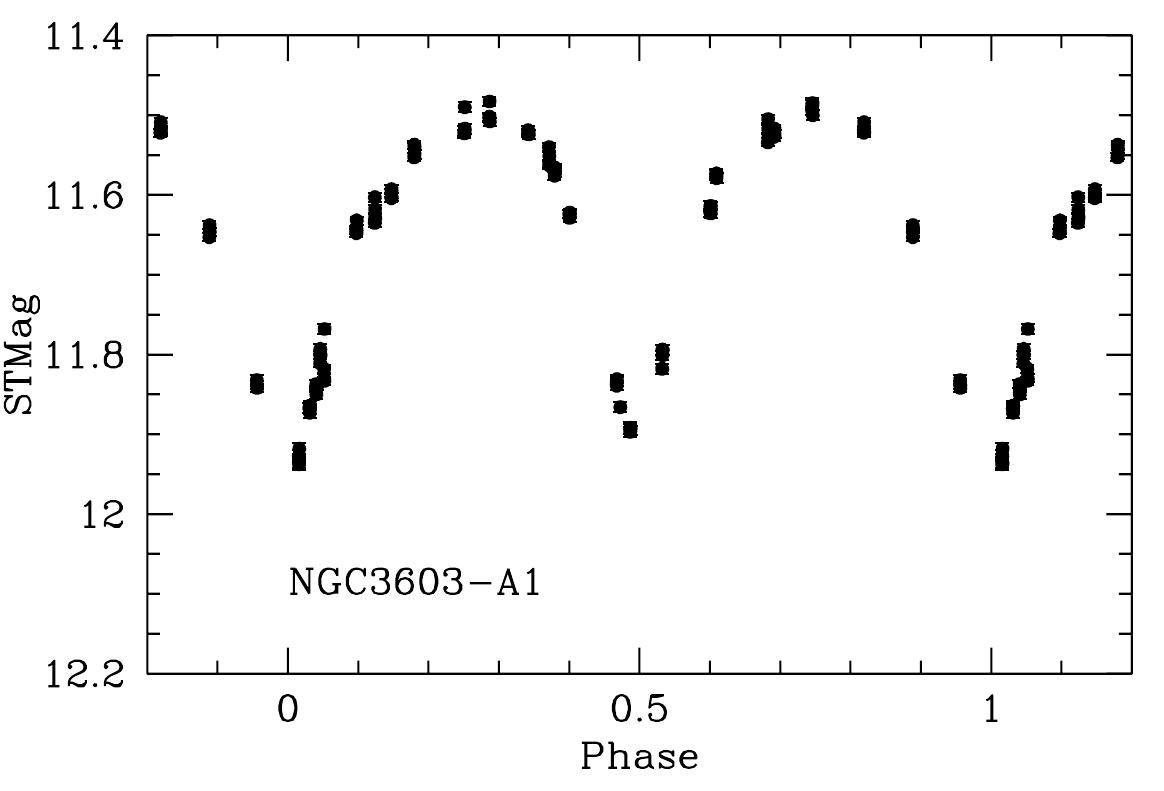}
\caption{\label{fig:phot} Phased photometry of NGC~3603-A1. The phases were computed based on the
ephemeris we derive here, P=3.77298 days and T0=2456161.365.  Error bars are included
but are often masked by the point size.}
\end{figure}

We ran a Lafler-Kinman period search \citep{1965ApJS...11..216L} on our STIS photometry and derived a revised period of 3.77298$\pm$0.00005 d. Plotting the
data allowed us to revise the phase zero point to T0=2456161.365$\pm0.03$.  
In Figure~\ref{fig:phot} we show the phased photometry.  The two eclipses are similar in depth, with the larger one   
about 0.45~mag deep, and the other about 0.40~mag deep. (The NICMOS J-band photometry of \citealt{2004AJ....128.2854M} may show
somewhat more shallow eclipse depths, but the their much larger photometric errors and their poor phase coverage at the middle of the eclipses  make it hard to draw strong conclusions.)  Note that the eclipses are separated by 0.5 phases, consistent with a circular orbit, as would be expected given the short period and large masses.

\subsection{Spectroscopy and Radial Velocities}

\label{Sec-spectra}

Determining the masses of the components of NGC~3603-A1 rests on our ability to measure the radial
velocities of representative spectral lines from each star.   
\citet{2008MNRAS.389L..38S} may have done the best that
could be done from ground-based observations, but were forced to measure radial velocities based on the locations
of the peaks of blended broad emission lines; see, e.g., Figure 2 in their paper.  The same team obtained 14 orbits
during Cycle 19 to obtain time-resolved optical spectroscopy of NGC~3603-A1 (GO-12615, PI: Schnurr) using 
the G430M/3936, G430M/4706, and G750M/7283 settings.   The orbits were timed for full phase
coverage.  In all, 12 orbits were devoted to these optical observations, spaced roughly at 0.1 phase intervals, with the other two (Visits 5 and 10) 
used for UV observations.  The optical spectra were taken using the 52\arcsec$\times$0\farcs2 slit, after offsetting from Sher 25, an isolated, nearby (20\arcsec) B1~Iab
star.
As mentioned above, four 0.1~s images were obtained at the start of each of the 14 visits.  Until now, none of these
data have been used in any analysis, other than a poster our group presented at a meeting \citep{2021AAS...23713305B}.

\begin{figure}
\epsscale{0.65}
\plotone{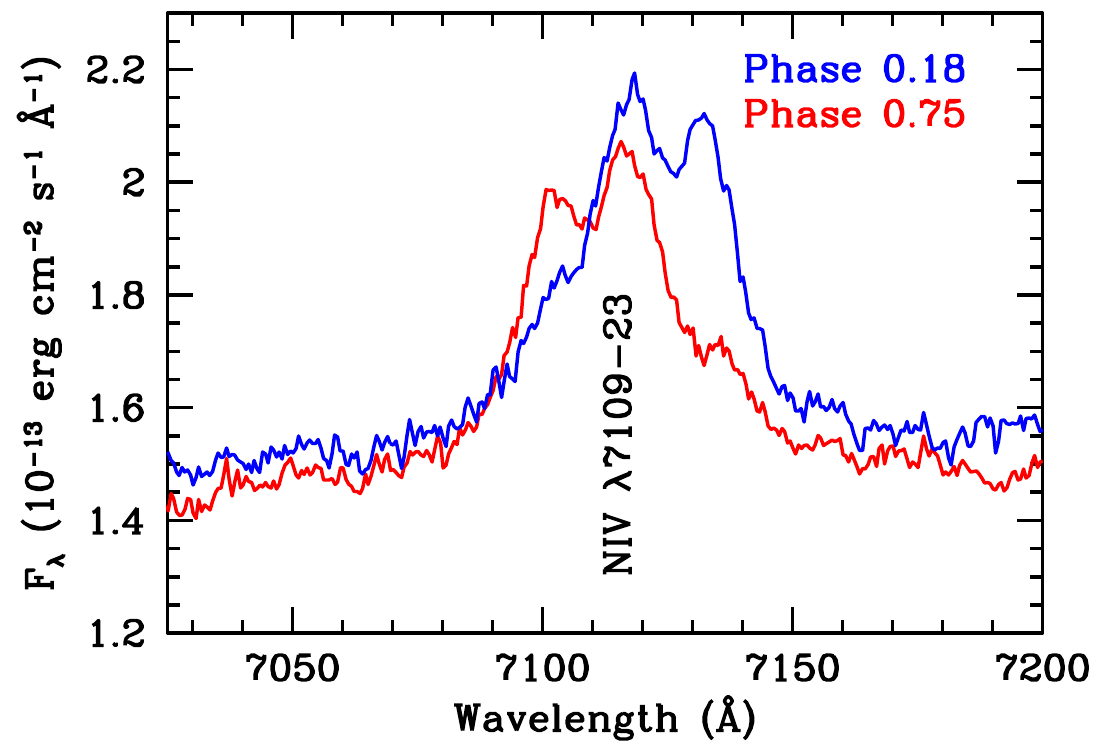}
\plotone{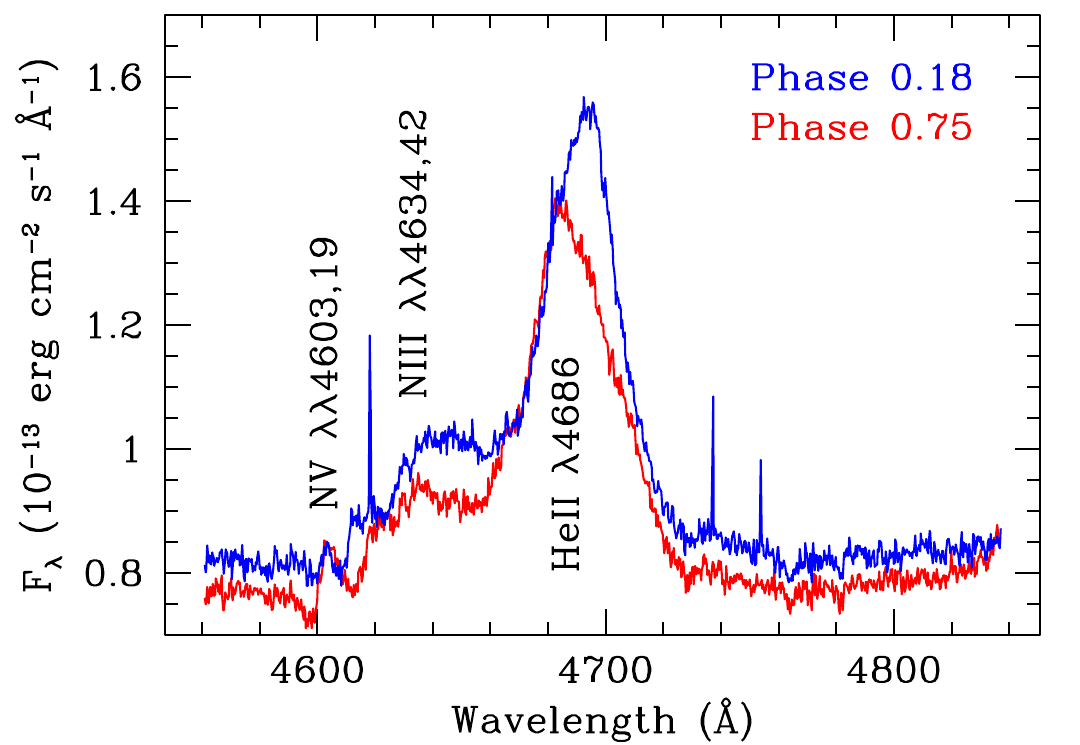}
\plotone{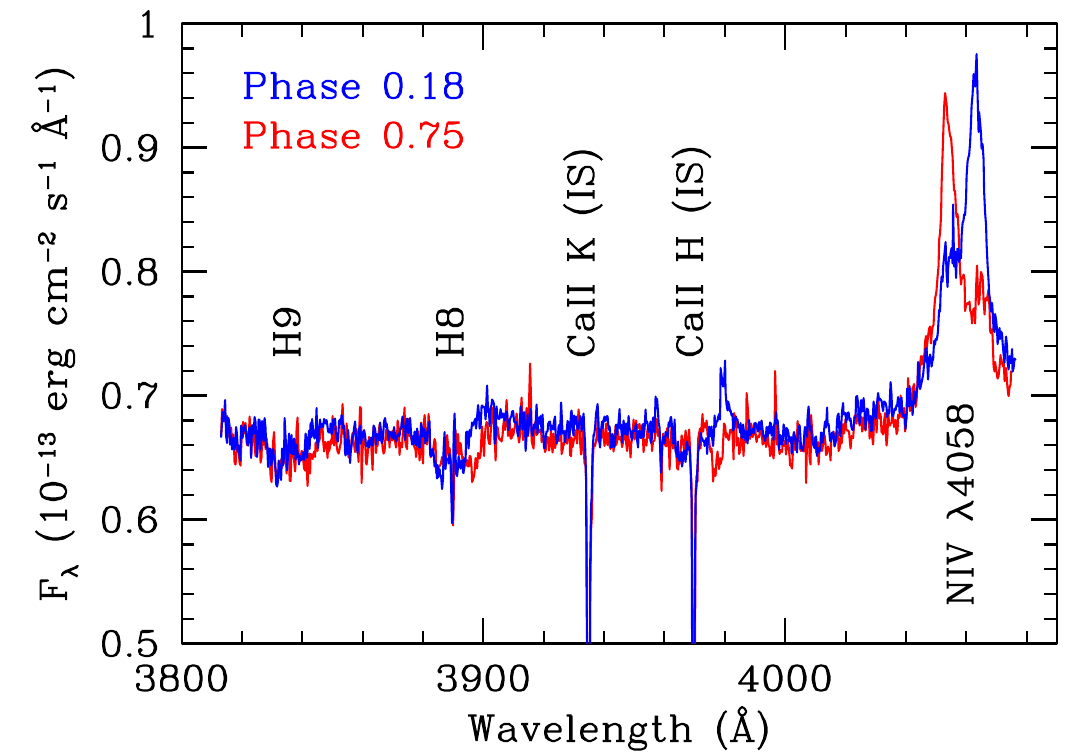}
\caption{\label{fig:blended} Phase shifts in the archival STIS spectra. To demonstrate the largest change with orbital phase, we
show the spectra from 12615/V4 (phase 0.18, blue) and 12615/V12 (phase 0.75, red).  The top panel shows a zoomed image
of the N\,{\sc iv} 7109-23 complex, the only line visible in the G750M/7283 exposures.  The apparent double peak is not due to the primary and secondary, but rather to the two principle components of this line, as we will show later (Sec.~\ref{Sec-dis}).  The middle panel shows the entire
coverage of the G430M/4706 spectrum.   There is a strong shift in the shape of the He\,{\sc ii} $\lambda 4686$ line, along with shifts with the N\,{\sc v} $\lambda \lambda$ 4603,19 and N\,{\sc iii} $\lambda \lambda$ 4634,42 doublets.  In the bottom panel (taken with the G430M/3936 setting), the N\,{\sc iv} $\lambda 4058$ line is double peaked, with the stronger and weaker components exchanging places with phase.  None of these features lend themselves to measuring reliable radial velocities due to the extreme blending.  However,
a careful inspection of the two absorption features, H8 and H9, shows that they are double.  This discovery provided a path forward for this project.  The data shown in this figure all come from program ID 12615 (PI: Schnurr), and were re-processed here.}
\end{figure}

We examined these spectra, and our first impression was that we could understand why nothing had come of them: most showed only asymmetrical lines that changed shape from observation to observation.  Examples are shown in Figure~\ref{fig:blended}. 
 However, co-author S.B. very astutely noted that on several of the G430M/3936 exposures the upper Balmer lines, in particular H8 and H9 
(present in absorption, not emission), were clearly double. (See the bottom panel in Figure~\ref{fig:blended}.)   In addition to providing a much cleaner approach than trying to
deconvolve blended, broad emission lines, such absorption lines are more likely to represent the true motion of the star than emission lines
formed in a stellar wind (see, e.g., \citealt{1979ApJ...228..206C,2008MNRAS.389.1447N,2024A&A...689A.157S}).

This inspired us to propose for additional HST/STIS spectra.  After several tries, we were successful in Cycle 31 (GO-17527, PI: Massey).  We had learned from the archival data that there was little point in
obtaining data at phases other than those near orbital quadratures  when the line separations would be the largest.  We therefore asked for four orbits, to permit two independent observations near phase 0.25 and two near phase 0.75, using our revised ephemeris (Section~\ref{Sec-photometry}). We chose the G430M/3843 grating setting as this would provide 50 km s$^{-1}$ spectral resolution and cover the 3700-3980\AA\ region, with H9 well centered.  
  To deal with the hot pixels and
cosmic rays, we dithered along the slit to 7 positions, each separated by 0\farcs5 ($\sim$10~pixels).  At each position we performed two exposures (CR-SPLIT=2) in order to reduce cosmic rays.  Thus each observation consisted of 14 individual exposures taken at 7 positions, each of 155 s.  This proved extremely effective in eliminating bad data. 

Crowding meant that we  had to be careful to avoid stars falling within the slit near A1, and so we restricted the allowed orientation angles in designing our Phase II. 
  We used the 52$\times$0.1E1 aperture, which places the target near the top of the
array (y$\sim$900) in order to minimize the CTE losses.  The use of the narrow slit (0\farcs1) allowed us to permit a wider range of orientation angles without having contamination by other stars.   The dithering resulted in a final position at y=960, still well away from the edge at y=1024.

In order to facilitate target acquisition, we refined the positions of both the offset star (Sher 25) and A1 from archival images.  Our best
values are $\alpha$=11:15:07.626, $\delta$=$-61$:15:17.56 for Sher 25, and $\alpha$=11:15:07.277, $\delta$=$-61$:15:38.39 for A1.
(Crowding compromises the Gaia positions, with the ``renormalized unit weight error" value for A1 given as 14.  Values
higher than 1.5 are considered compromised.)
After the offset from Sher 25, we performed an ACQ/PEAK to assure good centering in our narrow slit.
We are grateful to our program scientist, Dr.\ Alex Fullerton, for his considerable efforts working with us to refine both the position
and orientation angles.

Throughout the past few years, there has been increasing issues with guide star failures, and our program was no exception.
Visit 1 was successful. Visit 2 failed, and had to be rescheduled as Visit 5, which was successful.  Visit 3 was successful, but Visit 4 failed and had to be rescheduled as Visit 6, which also failed, and was rescheduled as Visit 7, which succeeded. 

For each of the 4 visits, the pipeline reduction produced an extracted  heliocentric wavelength- and flux-calibrated fits table for each of the 7 dithers. 
These have already had the CR-SPLIT pair of images combined in such a way to reject cosmic rays, leaving primarily hot pixels
to be removed by filtering.   We did this by taking the median of the 7 spectra, and confirming that the final signal-to-noise was as expected.

A larger challenge was dealing with the archival spectra.  These had not been CR-SPLIT, and they had been dithered to only 3 positions. Inspection showed that the redwards component of the H9 line was strongly affected by bad pixels at the third dither
location.  Taking the median of those three exposures worked as well as anything we tried, and was adequate.

\subsubsection{Radial Velocities}

\begin{figure}
\epsscale{0.7}
\plotone{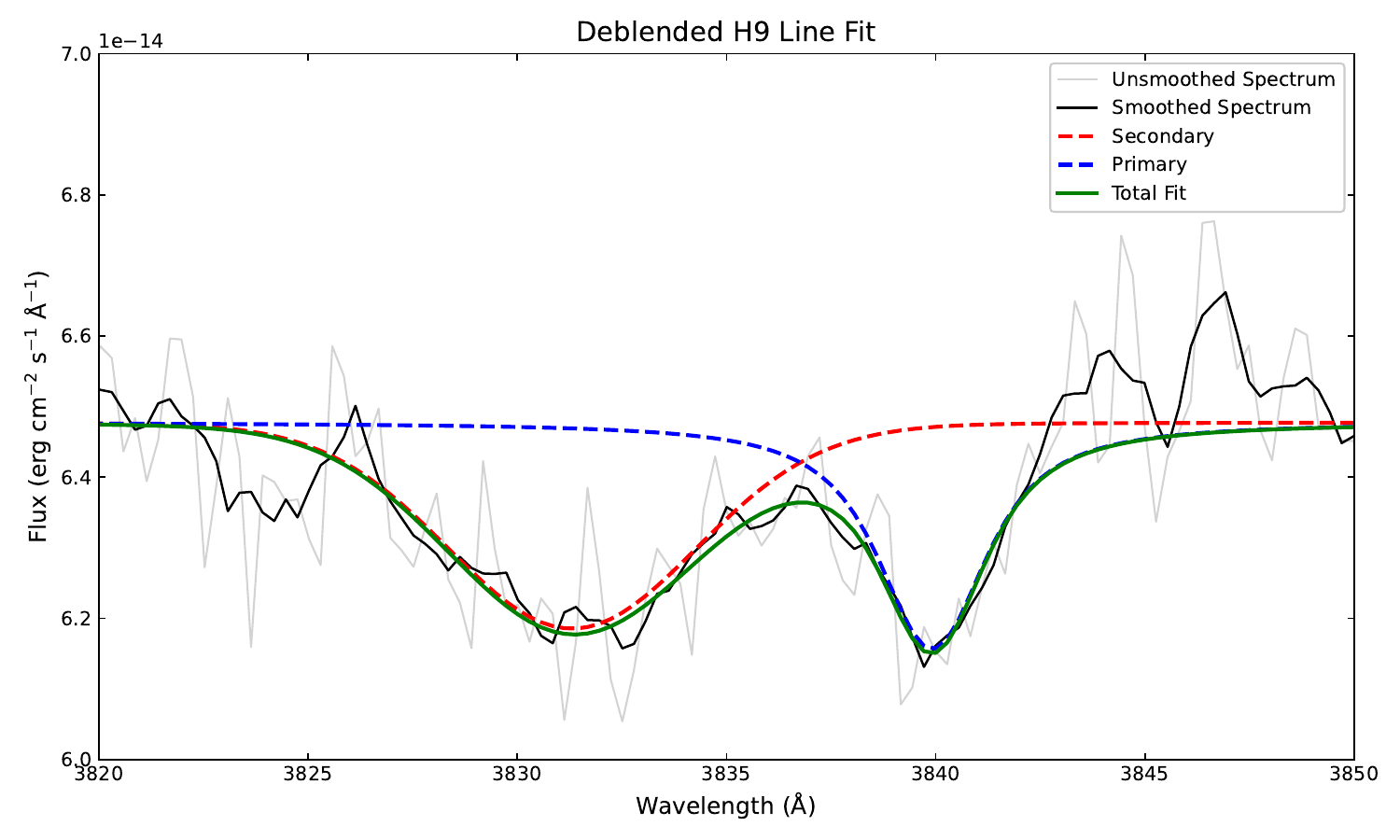}
\caption{\label{fig:example} Example of a fit to the H9 line.  The light grey solid curve shows the unsmoothed data; the solid black curve shows the spectrum smoothed by a 5-point boxcar.  The data are from GO-17527/V1. The component of the line on the left is from the secondary, while that on the right is
from the primary.  The ``deblend" routine in IRAF's {\sc splot} was used to fit Voigt profiles to the two components.  The individual fits are shown
by dashed lines, red for the secondary, and blue for the primary.  The solid green line shows the composite fit,  a good match to the observation. Note that the secondary's profile is much broader than the primary's, indicating a larger rotational velocity.}
\end{figure}

The H9 upper Balmer line has a laboratory (vacuum) value of 3836.472~\AA. The two components are well separated on the
data from our four successful visits (GO-17527/V1, V3, V5, and V7), as expected given their timing at quadrature.  We measured their wavelengths using
the versatile deblend option in IRAF's {\sc splot}, utilizing Voigt profiles, fitting both components
simultaneously. We measured each spectrum five times to get some sense of the internal precision, using 
3-7 pixel boxcar smoothing.  An example is shown in Figure~\ref{fig:example}.  We find that the secondary has a much broader profile than that of the
primary; we estimate projected rotational velocities $v\sin{i}$ of $500\pm45$ km s$^{-1}$ for the secondary, and $280\pm40$ km s$^{-1}$ for the primary.

For the archival data, only four visits showed double lines, as expected from their phases (GO-12615/V4, V11, V12, and V13).  
We measured the lines in the same manner.  The resulting heliocentric radial velocities are given in Table~\ref{tab:RVs}.

The $\sigma$ values only represent our internal precision.  There are many random and systematics involved, and we
estimate that the true uncertainties involved are likely 30 km~s$^{-1}$.   We also note that although H8 was seen as double, it was
affected by cosmetic problems on both sets of data and not measured.  The weights listed in Table~\ref{tab:RVs} are based on a visual impression of
the spectra, basically either ``good" (assigned a weight of 1.0) or ``poor" (assigned a weight of 0.5).

\begin{deluxetable}{l l l r c r r c c} 
\tablecaption{\label{tab:RVs} Radial Velocities NGC~3603-A1}
\tablehead{
\colhead{ID/Visit}
&\colhead{HJD}
&\colhead{Phase}
&\multicolumn{2}{c}{Primary (km s$^{-1}$)}
&
&\multicolumn{2}{c}{Secondary (km s$^{-1}$)}
&\colhead{Weight} \\ \cline{4-5} \cline{7-8}
&&&\colhead{RV} &\colhead{$\sigma$}
&&\colhead{RV} 
&\colhead{$\sigma$} 
}
\startdata
12615/V13&2456096.5558 &0.823  &    -257.6 &     0.7&& 355.2   &   4.4 &1.0 \\
12615/V11&2456099.8144 &0.686  &     -367.1 &     1.0&& 436.0   &   1.1 &1.0 \\
12615/V12&2456100.0520 &0.749  &     -321.6 &     4.0&&414.5   &   0.4 &0.5 \\
12615/V4 &2456101.6904 &0.184  &     230.4 &     3.1&&-369.5   &   1.3 &1.0 \\
17525/V1 &2460561.5641 &0.240  &     271.9 &     1.0&&-385.9   &   2.5 &1.0 \\
17527/V3 &2460563.4032 &0.727  &    -419.2 &     1.6&& 396.3   &   1.5 &1.0 \\
17527/V5 &2460629.6184 &0.277  &   217.9 &     0.9&&-408.5   &   1.9 &1.0 \\
17527/V7 &2460733.3415 &0.768  &    -323.7 &     1.5&&314.9   &   1.8 &0.5 \\
\enddata
\end{deluxetable}

\section{Analysis}
\label{Sec-analysis}

An orbit solution based on the data in the previous section will be a significant
improvement, but it will still be dependent on only eight observations and using a single
spectral line.  We therefore want to test the robustness in as many ways as possible.

\citet{1941ApJ....93...29W} pointed out that the {\it mass ratio} could be readily determined for
a spectroscopic binary without knowledge of the actual orbital parameters by simply plotting the
radial velocities of each component against each other.  The points should fall along a straight
line, with a slope  $r$, which is the negative of the inverse of the mass ratio, i.e., 
$$r = \frac{\Delta {\rm RV_{\rm pri}}}{\Delta {\rm RV_{\rm sec}}}=-\frac{K_{\rm pri}}{K_{\rm sec}}=-\frac{m_{\rm sec}}{m_{\rm pri}}=-q$$
where m is the mass and $K$ is the orbital semi-amplitude.  
The advantage of this method is that it provides a robust answer for a scant number of points, and gives a value that is independent of any assumptions of the period or orbital eccentricity.  We employ it to provide
a reality check on our orbital parameters.

\begin{figure}
\plotone{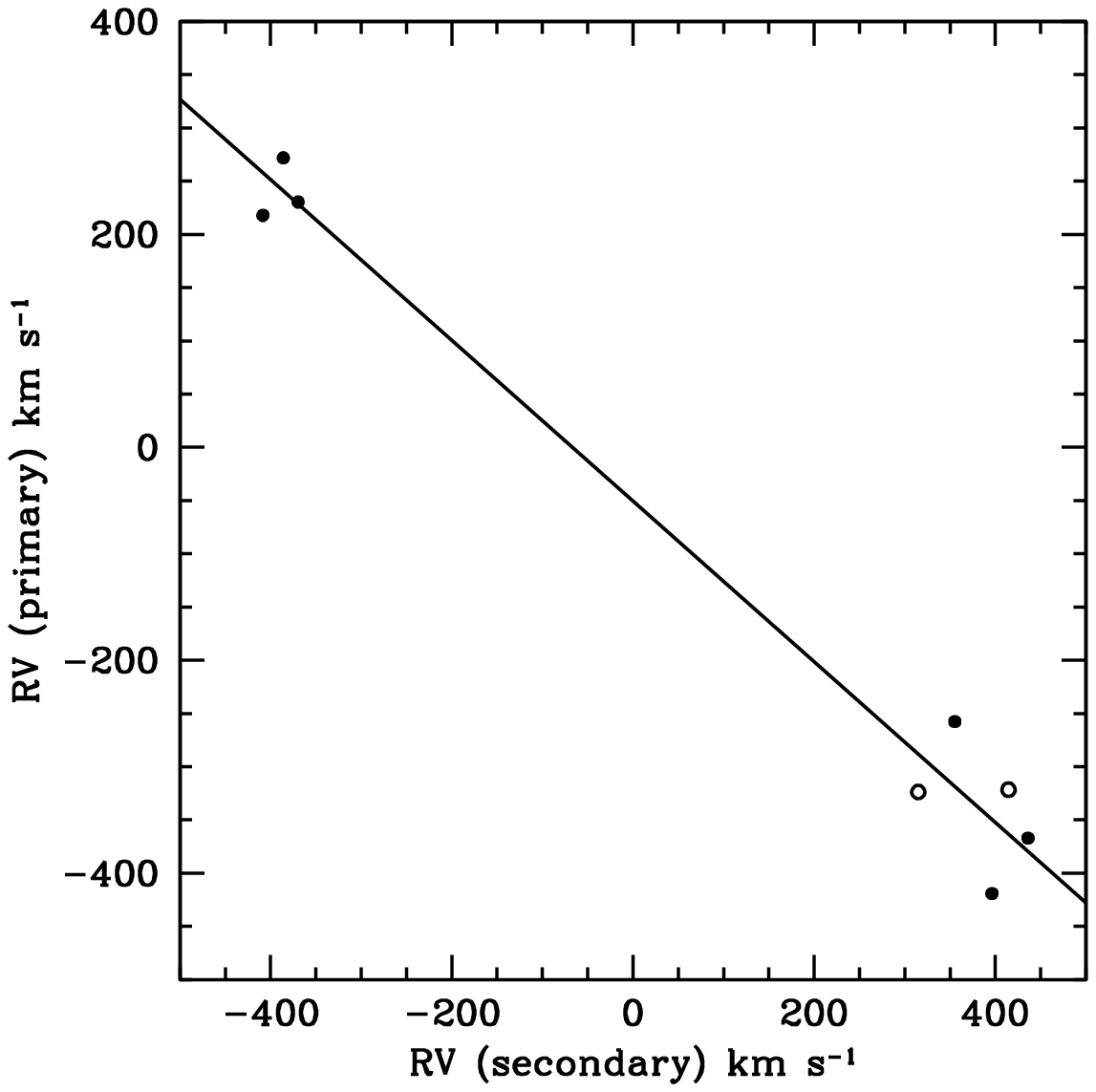}
\caption{\label{fig:wilson} A ``Wilson Diagram" for NGC 3603-A1.  The radial velocities of the primary star are plotted against the radial velocities of the secondary star.  The two points shown as open circles are the ones with lower weight given in Table~\ref{tab:RVs}.  The -0.75 slope implies a mass of the primary  1.3$\times$ that of the secondary star.}
\end{figure}

We show such a Wilson diagram in Figure~\ref{fig:wilson}.  We calculated the straight line fit by minimizing the orthogonal distance from each
point.  This provides a better estimate than conventional least-squares methods which assume no uncertainty in the abscissa and merely minimizes
the deviation in the y direction.  We find a slope $r=-0.755\pm0.045$ if we assign weights as given in the final column of Table~\ref{tab:RVs}, and
a slope $r=-0.753\pm0.046$ if we assign equal weights.  Weighing the points by $1/\sigma^2$ results in a slope $r=-0.756\pm0.041$.  Thus, the mass
ratio, $q$ (taken to be $-r$, or $m_{\rm sec}/m_{\rm pri}$), is quite insensitive to the weighting options used.   This value is consistent with the preliminary $q=0.72\pm0.03$
value reported earlier \citep{2021AAS...23713305B}  based on more limited data.   The orbital semi-amplitudes derived by
\citet{2008MNRAS.389L..38S} imply a mass ratio  of $0.76\pm0.10$.

We note that we are flouting the usual convention that the deeper photometric minimum defines the primary star as the one being eclipsed  \citep{2001icbs.book.....H}.  For binaries whose components are unevolved, this star will be the more massive one, since the deeper eclipse will correspond to the hotter star being eclipsed. For evolved systems, this is not necessarily the case, i.e., the hotter star might be the less massive one.  Our zero-point definition is the same (within N complete cycles) as that used by all previous studies of NGC~3603-A1 (\citealt{1984ApJ...284..631M, 1985ApJ...295..109M,2004AJ....128.2854M}, and
\citealt{2008MNRAS.389L..38S}), i.e., with the more massive star in inferior rather than superior conjunction at phase 0.  Thus to avoid confusion we will continue to refer to the ``primary" star as being the more massive one.  The primary star is also the one with the stronger
emission lines.  This was also found by \citet{2008MNRAS.389L..38S} in their  NIR spectra, and we see this clearly in the behavior of the emission
lines by phase shown in Figure~\ref{fig:blended}.

\begin{figure}[h]
\epsscale{0.8}
\plotone{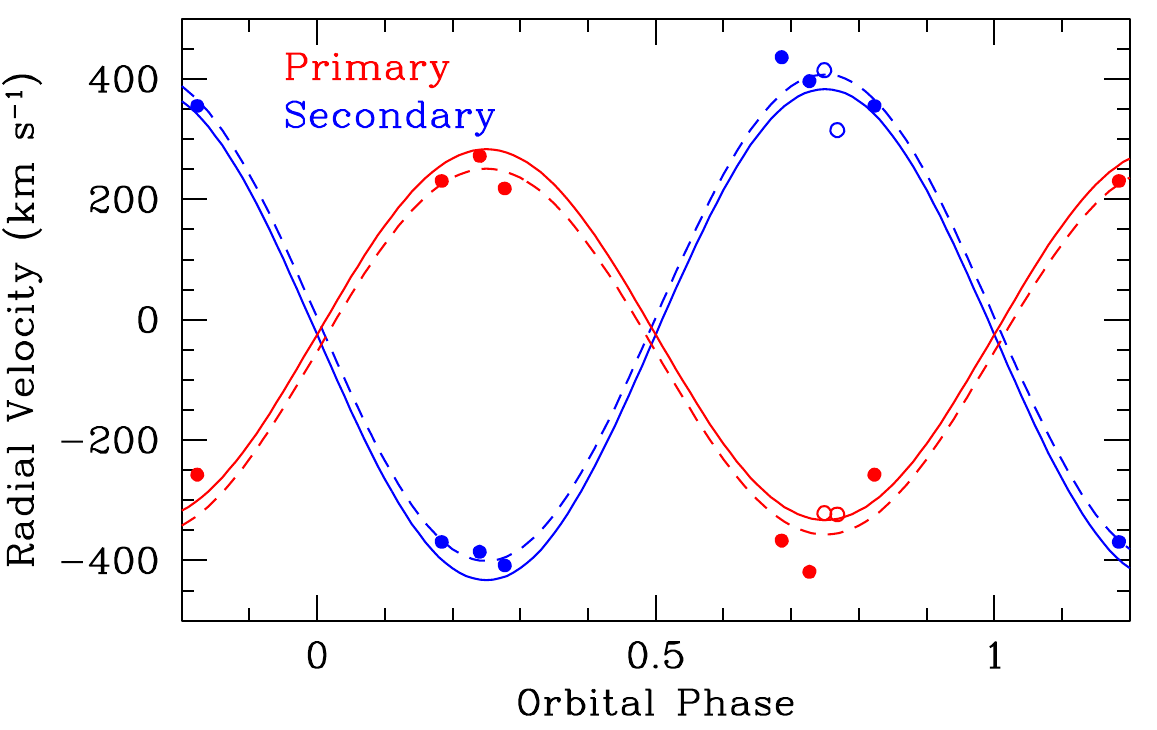}
\caption{\label{fig:orbit} Orbit solution for NGC~3603-A1. The solid lines show the orbital solutions given in Table~\ref{tab:orbit}.
for $\gamma$ fixed to the same value for both the primary and secondary.  The dotted lines show the solutions for individual values of $\gamma$.
The two open circles show the two half-weighted pairs of points.}
\end{figure}

We next use our radial velocities to determine orbit solutions for both components. Given the short period and large masses we
expect the orbits to be circular, and this is consistent with what we see by eye in Figure~\ref{fig:phot}, that  the primary and secondary eclipses are separated by 0.5 phase. We give the orbital parameters in Table~\ref{tab:orbit} along with the root-mean-squares (RMS) of the fits and the minimum masses of the two components, and show the orbit solution by the solid curves in Figure~\ref{fig:orbit}.

\begin{deluxetable}{l c  c} 
\tablecaption{\label{tab:orbit} Orbital Fits for NGC~3603-A1}
\tablehead{
\colhead{Parameter}
&\colhead{Primary}
&\colhead{Secondary}
}
\startdata
\cutinhead{Single $\gamma$} 
$P$ (days) & \multicolumn{2}{c}{$3.77298\pm0.00005$} \\
$T0$ (HJD) & \multicolumn{2}{c}{$2456161.365\pm0.03$} \\
$e$ (fixed) &\multicolumn{2}{c}{0.00} \\
$\gamma$ (km s$^{-1}$) & \multicolumn{2}{c}{$-24.9\pm13.3$}\\
$K$ (km s$^{-1}$) & $307.9\pm19.6$ & $408.2\pm19.6$  \\
Mass ratio $q$ (Sec/Pri)& \multicolumn{2}{c}{$0.754\pm0.060$} \\
RMS (km s$^{-1}$)& 41.5 & 39.6  \\
$a\sin{i}$ ($10^6$ km) & $16.0\pm1.1$& $21.1\pm1.0$ \\ 
$m\sin^3{i}$ ($M_\odot$)& $82.2\pm9.6$ & $62.0\pm8.1$ \\
\cutinhead{Separate $\gamma$}
$P$ (days) & \multicolumn{2}{c}{$3.77298\pm0.00005$} \\
$T0$ (HJD) & \multicolumn{2}{c}{$2456161.365\pm0.03$} \\
$e$ (fixed) &\multicolumn{2}{c}{0.00} \\
$\gamma$ (km s$^{-1}$)& $-53.1\pm17.4$ & $3.3\pm14.2$  \\
$K$ (km s$^{-1}$)  & $304.0\pm18.2$ & $404.2\pm14.8$   \\
Mass ratio $q$& \multicolumn{2}{c}{$0.752\pm0.053$} \\
RMS (km s$^{-1}$) & 41.1   & 39.0  \\
$a\sin{i}$ ($10^6$ km) & $15.8\pm0.9$& $21.0\pm0.7$  \\
$m\sin^3{i}$ ($M_\odot$) & $79.5\pm7.7$ & $60.1\pm7.1$  \\
\enddata
\end{deluxetable}

In computing the minimum masses and projected orbital separations,  we used Monte Carlo sampling in order to account for their non-linear dependence on $P$, $K_1$ and $K_2$.  The Monte Carlo approach allows us to sample the full posterior distributions of the inputs, thus capturing the true asymmetric and skewed uncertainties of the derived quantities.  This is particularly important because the mass function depends linearly on the period but cubically on the velocity
amplitudes, making the use of linear approximations inadequate.  Our method yields slightly higher mean masses than the traditional estimate (by a few tenths of a solar mass), but reflects a more accurate integration over the observational error distributions.

For most binaries one can expect the center-of-mass velocity ($\gamma$) to be identical for the two components.  This is not
necessarily the case for very massive, luminous systems, as even absorption lines can be formed at a point in the atmosphere where
there is a significant outflow from the stellar wind.  This was demonstrated by studies in single stars, which showed different
velocities depending upon the ion \citep{1968MNRAS.141..219H} and confirmed statistically \citep{1977ApJ...214..759C}. Soon afterwards, analysis of two Of/WN ``transition" binaries, HDE~228766 and BD+40$^\circ$4220, showed differences in the $\gamma$ velocities derived for
the primary and secondary stars \citep{1977ApJ...218..431M}.  (More recent examples of the need for this methodology can be found in  \citealt{2002ApJ...565..982M} and \citealt{2014ApJ...789..139M}.)  We therefore have also computed
the orbital parameters for NGC~3603-A1 without requiring that the $\gamma$ velocities be the same. 
The orbital fit with $\gamma$ as a free parameter for each star is shown by dashed curves in Figure~\ref{fig:orbit}.

We note explicitly that we were able to determine reliable orbit solutions with only eight data pairs thanks to being able to adopt the phasing determined by the light curve
ephemeris.  We have adopted a similar procedure in previous studies (e.g., \citealt{2012ApJ...748...96M}).  We are encouraged by the fact that the ratios of the derived semi-amplitudes ($0.754\pm0.060$, single $\gamma$; $0.752\pm0.053$ separate $\gamma$s) 
agree well with those determined from the Wilson diagram, $0.755\pm0.045$ (weighted) and $0.753\pm0.046$ (no weights). Although the data are the same,
the methodology is radically different, as the Wilson diagram makes no assumption about orbital phases or orbital eccentricities.

In the end, we adopt the solution with the same $\gamma$ velocities, as the residuals from the curves are not significantly improved by
adding the additional degree of freedom.

\subsection{Orbital inclination and physical parameters from the light curve}
\label{Sec-laura}

Analysis of the light curve allows us to determine the orbital inclination; it also provides good estimates of other stellar parameters,
such as effective temperatures and stellar radii. 
To accomplish this, we used the light curve synthesis code {\sc gensyn} \citep{1972MNRAS.156...51M}, constraining the fit using the data from
the orbit solution given in the previous section.  

The code produced model differential light curves in STMag, V, and B bandpasses based upon trial input parameters.  We began with the expectation that the
effective temperature of the components would be in the 40,000-44,000~K range, based on the \citet{2010MNRAS.408..731C} analysis
of several O3If*/WN6 stars, including the (combined) spectrum of the NGC~3603-A1 system. 

We estimate the physical fluxes and limb darkening coefficients in each band from values from \citet{1979ApJS...40....1K} and \citet{1985A&AS...60..471W}, respectively. The physical fluxes were chosen at the central wavelength of each band: STMag at 5570~\AA, V at 5470~\AA, and B at 4330~\AA. We found no discernible distinctions between our STMag- and V-band differential light curves. Each trial run of {\sc gensyn} is set by six independent parameters: the effective temperatures of the components, the system inclination $i$, the mass ratio of the binary $q$, and the primary and secondary filling factors. {\sc Gensyn} defines the filling factor to be the ratio of the photospheric surface to the Roche surface. In addition to the unequal eclipse depths, the observed light curve displays significant ellipsoidal variations, indicating that at least one of the components is close to filling its Roche surface. The radii of the two stars are set from the semi-amplitude value of the primary, the mass ratio, and the filling factors.

For each run, we attempt to match three observables: the eclipse depths, the eclipse widths, and the ellipsoidal variations. The ratio of eclipse depths is highly dependent upon the temperature difference between the two stars. The radii are constrained by the eclipse widths and the ellipsoidal variations. And finally, the depth of the eclipses constrains the inclination.

It will be useful to compare the absolute visual magnitude derived from the light curve analysis to that derived from the star's photometry.
\citet{2008AJ....135..878M} find a spectroscopic parallax to the cluster of 7.6~kpc,  and note that this is also consistent with the kinematic
distance.  Their photometry (determined from HST/ACS data) has $V=11.18$ and $B-V=1.03$.   
The HJD of the ACS observation is 2453734.42, which corresponds to an orbital phase of 0.77; i.e., the data were taken outside of eclipse. Analysis of the photometry and spectroscopy
of neighboring stars (P. Massey et al.\ 2025, in prep) establishes that $E(B-V) = 1.25$ is a good approximation. \citet{2000PASJ...52..847P} investigated the extinction seen towards the cluster, and suggested a two-component model, with $$E(B-V)_{\rm total} = E(B-V)_{\rm foreground} + E(B-V)_{\rm {internal}}.$$  They assumed a normal value of 3.1 for the ratio of total to selective extinction, $R_V,$ applied to the foreground
reddening, but found a higher value, $R_V=4.3,$ applied to the additional reddening internal to the cluster.  Thus,  
$$A_V=3.1 \times 1.1 + 4.3  \times [E(B-V)-1.1].$$ Applying this, and correcting for a distance of 7.6~kpc, results in an absolute visual magnitude $M_V$ of the combined A1 system of $-7.3$.

 Our best fit models have an inclination of 73\fdg5.  (As we show below, we find we can exclude inclinations below 71$^\circ$ and above 76$^\circ$.) All good models show that the system is {\it almost} in contact, but whether the primary or secondary is filling their Roche surfaces is uncertain.  
 
\begin{figure}[h]
\epsscale{0.8}
\plotone{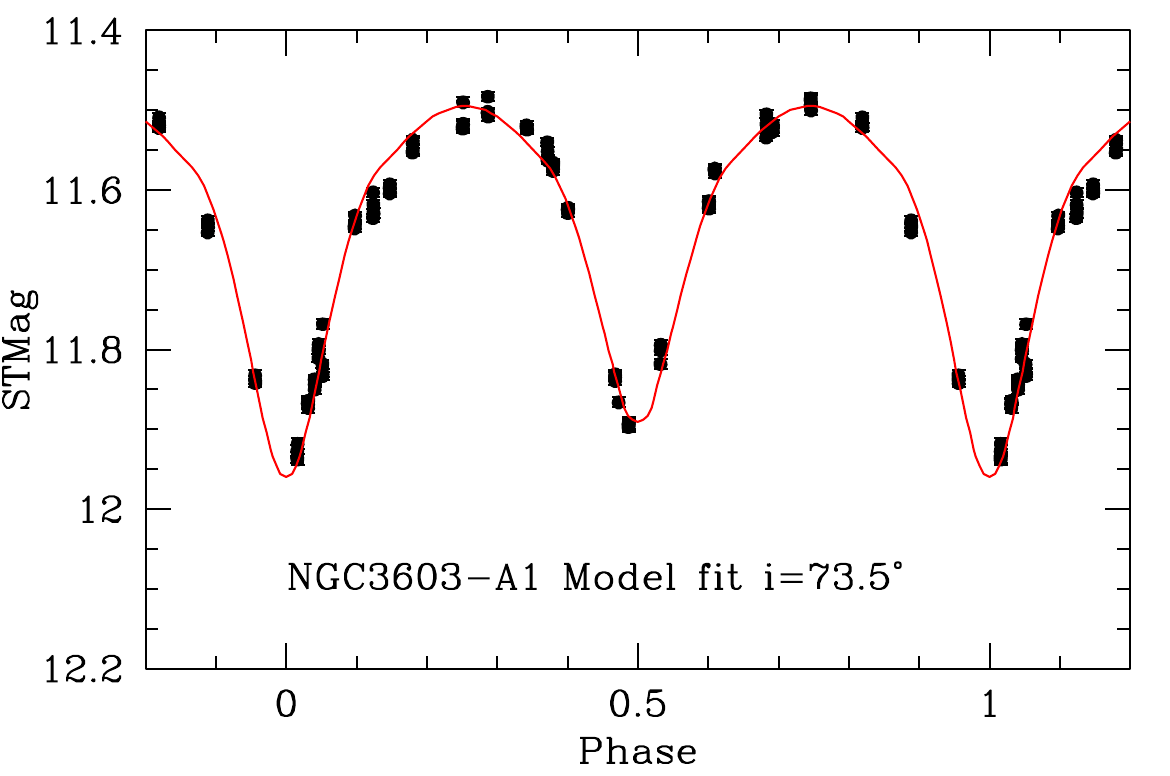}
\caption{\label{fig:LP} light curve model.  Our phased photometry is shown as points, with the red line showing
the best fit light curve.}
\end{figure}

The fit shown in Figure~\ref{fig:LP} has filling factors of 1.00 and 0.96 for the primary and
secondary, respectively.  We were unable to fit the eclipse depths for inclinations smaller than 71$^\circ$. At this inclination, the secondary is also filling its Roche surface. This model has the largest radii for both stars, $R_{\rm pri} = 22.9 R_\odot$ and $R_{\rm sec} = 20.1 R_\odot$, and the largest flux ratio of 0.901. Models at inclinations larger than 76$^\circ$ require the secondary to be unnaturally small to match the eclipse depths. Those models also do not accurately produce the observed ellipsoidal variations. For an inclination of 76$^\circ$, our model has primary/secondary radii of 22.3/14.9 $R_\odot$, and our smallest flux ratio is 0.518. All acceptable models with the primary filling its Roche surface have the primary star 2,000-5,000K cooler than the secondary. We adopted a $T_{\rm eff}$ = 42,000~K for the secondary, guided in part by \citet{2010MNRAS.408..731C}.  However, we found we could produce model fits with secondary $T_{\rm eff}$ from 38,000~K to 50,000~K, which result in $M_V$ (total) ranging from $-7.2$ to $-7.4$. The value of the orbital inclination is quite insensitive to the adopted effective temperature.  Note that our expectation is that $M_V=-7.3$, as discussed above.  We explored the effect of changing the orbital semi-amplitudes (i.e., the mass ratios) and found the derived inclination was also quite insensitive to those as well.  The final derived parameters are given in Table~\ref{tab:orbitfin}.
In computing the masses corrected for the inclination, we again used a Monte Carlo calculation, but using a Gaussian distribution centered at 73\fdg5 but truncated at 71\fdg0 ad 76\fdg0.

 We did also found a good fit to the light curve if we assumed that the secondary, and not the primary, was filling its Roche surface. That
 solution had filling factors of 0.92 and 1.00 for the primary and secondary, respectively, and 
 yielded similar temperatures.   The radius of the primary is smaller (19.8$R_\odot$) and the secondary larger (19.8$R_\odot$).  The flux ratio at 5570~\AA\ is 1.22, and the combined $M_V$ is $-7.2$, also a good match to our expectations.  
We found the same limits for the orbital inclination, with similar problems encountered at or below 71\fdg0 and at or above 76\fdg0, and yielded
the best fit at the same 73\fdg5 inclination. 
In Section~\ref{Sec-evol} we give the arguments for preferring the fit with the other solution: the rotation speed of the secondary suggests
it has been spun up by accretion, and the flux ratio would result in an unrealistically low bolometric luminosity for the massive primary.

\citet{2004AJ....128.2854M} analyzed the NICMOS light curve and reported parameters as a function of the mass-ratio, which were poorly constrained at the time.  According to their Table~2, the expected orbital inclination would be lower than ours, about 71$^\circ$.  They found that the secondary filled its Roche surface, and the primary almost filling its Roche surface. (Note that they refer to the secondary as
``the O star" or Star 1, and the primary as ``the WR star", or Star 2.)   However, their solution has the temperatures of the two stars opposite
to what we find.  Their light curve is quite noisy compared to ours, and the deeper eclipse occurs at their phase 0.5; compare their Figures 5 and 6 with our Figure~\ref{fig:LP}.

\begin{deluxetable}{l c  c} 
\tablecaption{\label{tab:orbitfin} Adopted Solution NGC~3603-A1}
\tablehead{
\colhead{Parameter}
&\colhead{Primary}
&\colhead{Secondary}
}
\startdata
Spectral Types & O3If*/WN6 & O3If*/WN5 \\
$P$ (days) & \multicolumn{2}{c}{$3.77298\pm0.00005$} \\
$T0$ (HJD) & \multicolumn{2}{c}{$2456161.365\pm0.03$} \\
$e$ (fixed) &\multicolumn{2}{c}{0.00} \\
$\gamma$ (km s$^{-1}$) & \multicolumn{2}{c}{$-24.9\pm13.3$}\\
$K$ (km s$^{-1}$) & $307.9\pm19.6$ & $408.2\pm19.6$  \\
Mass ratio $q$ (Sec/Pri) & \multicolumn{2}{c}{$0.754\pm0.060$} \\
$a\sin{i}$ ($10^6$ km) & $16.0\pm1.0$& $21.2\pm1.0$ \\ 
$m\sin^3{i}$ ($M_\odot$)& $82.2\pm9.6$ & $62.0\pm8.1$ \\
$i$ & \multicolumn{2}{c}{$73.5^\circ \pm 2.5^\circ$} \\
$T_{\rm eff}$ & 37,000~K & 42,000~K \\
$R (R_\odot$) & 22.6 & 18.4 \\
$M_V$ (mag) & \multicolumn{2}{c}{$-7.3$} \\
Flux ratio (Sec/Pri) & \multicolumn{2}{c}{0.794} \\
Filling factors & 1.00 & 0.96 \\
$m$ ($M_\odot$)&$93.3\pm11.0$ & $70.4\pm9.3$\\
\enddata
\end{deluxetable}

\subsection{Disentangling the Spectra}
\label{Sec-dis}

Based on the changing asymmetries of the broad emission lines, and the presence of the upper Balmer lines in absorption, we can confidently classify each component of NGC 3603-A1 as an O2-3If*/WN5-6 ``slash" star.   However, we can now go a little bit further.  Although the complete phase coverage provided by the archival spectroscopy was not helpful for measuring
radial velocities, it does prove useful in performing spectral disentanglement.  To do this, we adopt the expected radial velocity of each component based upon the orbital semi-amplitudes and (combined) $\gamma$ velocity. This operation is particularly challenging when the spectra are dominated by broad emission lines and the Doppler shifts are a small fraction of the line widths, as is the case here (see, e.g., Figure~\ref{fig:blended}).  Normally one would incorporate the flux ratio
in such disentanglement, but the value derived from the light curve applies mostly to the continnum, while the ratio of fluxes in the emission
lines will be mostly dominated by the relative mass-loss rates.  
Nevertheless, the exercise proved of some benefit.  

\begin{figure}
\epsscale{0.65}
\plotone{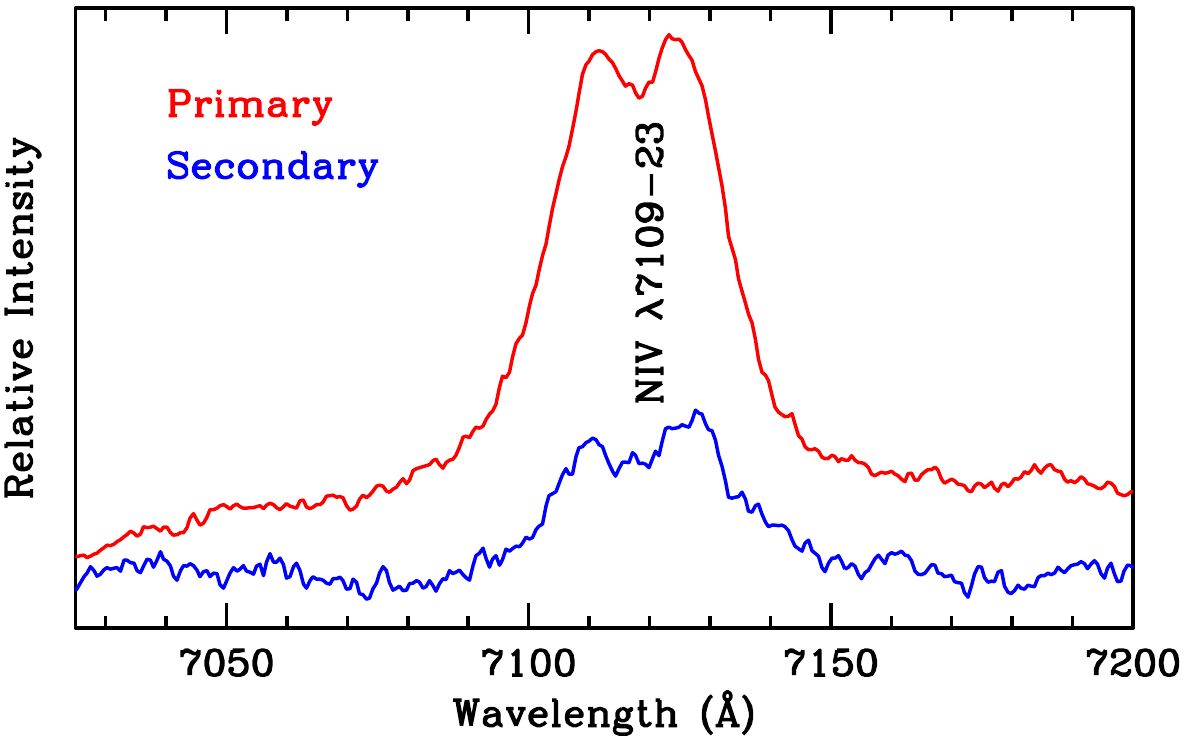}
\plotone{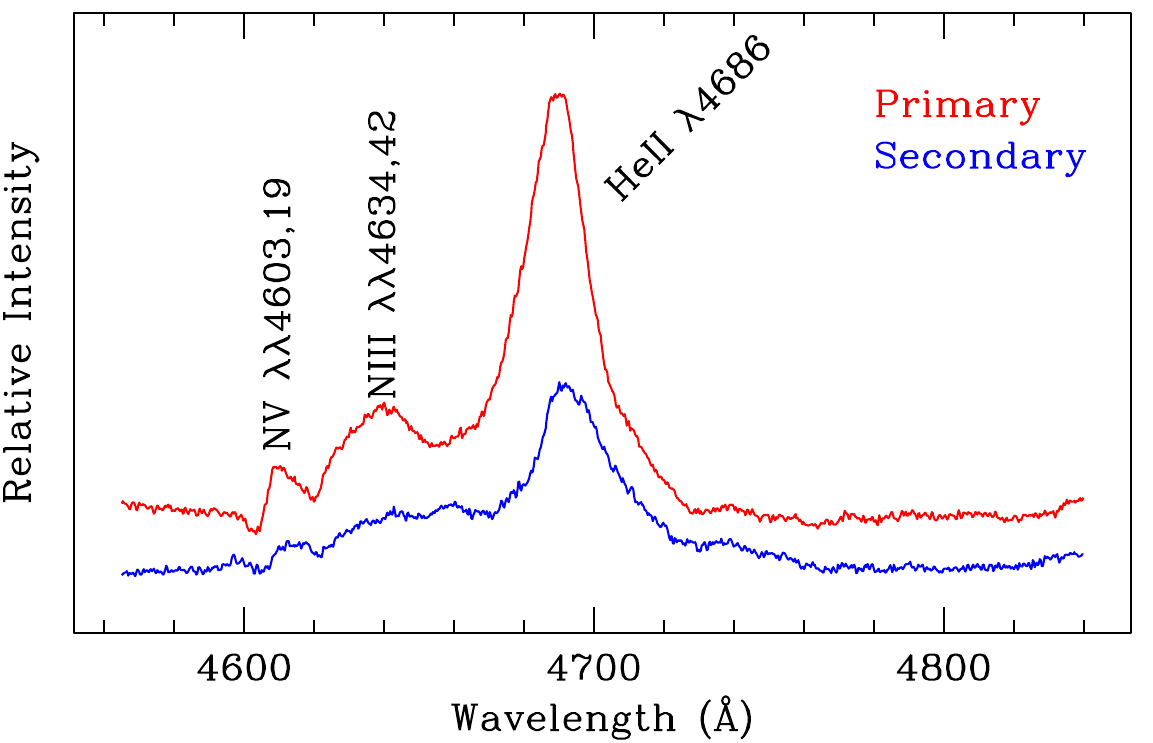}
\plotone{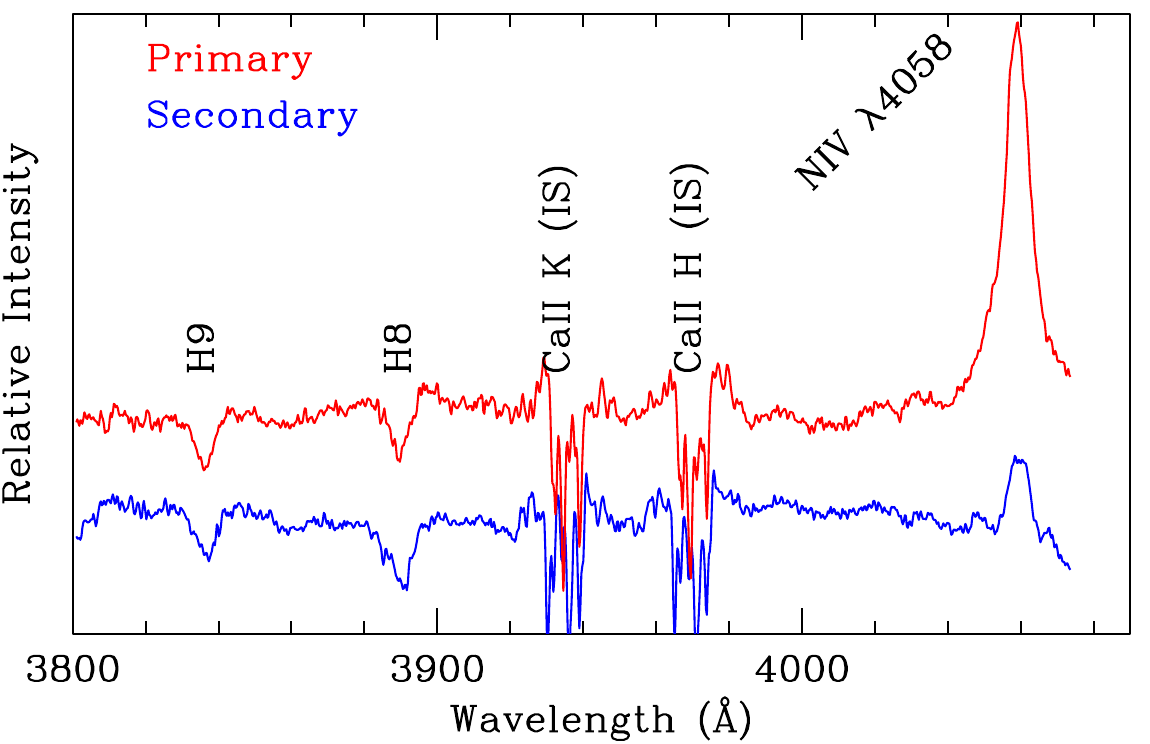}
\caption{\label{fig:dis} 
The disentangled spectra of NGC~3603-A1. The relative intensities should not be trusted, but this exercise confirms
that both components are O2-3I*/WN5-6 ``slash" stars.  The interstellar Ca\,{\sc ii} H and K  lines are smeared out since they
do not follow the radial velocity motion of either component. 
}
\end{figure}

We did this using a heavily modified version of the {\sc pydisentangle} Python package for spectral disentangling \citep{horton2025}, adapted and implemented with assistance from the ChatGPT AI model \citep{2023arXiv230308774O}.  

The resulting deconvolved spectra are shown in Figure~\ref{fig:dis}.  We do not learn a lot from this process, but
we do learn something.  The main thing we learn is that the primary and secondary both have similar spectral features, and
share the same emission lines.  Thus, calling them both O2-3If*/WN5-6 is not contradicted by the data. It is unfortunate that
features such as He\,{\sc ii} $\lambda 4200$ were not included in the archival spectroscopy, as we would expect this to show
both an emission and absorption component.  (See, e.g., Figure 3 in \citealt{Sota}.)

A comparison of the disentangled nitrogen features is also revealing.  
We see in the top panel that the two peaks in the N\,{\sc iv} $\lambda$7109-23 complex
are shared in both the primary and secondary spectra---each peak does not come from a single component. 
The middle panel shows that the primary has clearly defined N\,{\sc iii} somewhat stronger than N\,{\sc v}.
Given the strength of N\,{\sc iv} in the primary (bottom panel), we would call the primary's subtype WN6, not WN5.  However,
for the secondary, we find that while N\,{\sc v} is well recovered, N\,{\sc iii} is
not.  Thus, we are inclined to call the emission more typical of a WN5.   The higher excitation classification
is consistent with the secondary being the hotter star, as found in our analysis of the light curve (Section~\ref{Sec-laura}).

\section{Summary and Discussion}
\label{Sec-discussion}

We have performed a new analysis of the massive binary NGC~3603-A1, the brightest star in the NGC~3603 cluster.
Using both archival and newly obtained HST/STIS spectra, we have obtained new orbital elements indicating 
minimum masses of  $82.2\pm9.6$ and $62.0\pm8.1M_\odot$, with uncertainties much lower than
the previous analysis by \citet{2008MNRAS.389L..38S}.  We have also obtained a well-sampled, accurate light curve
using CTE-corrected photometry from STIS.  Constrained by the orbital data, the light curve yields an orbital inclination of 73\fdg5$\pm2.5$,
leading to masses of $93.3\pm11.0 M_\odot$ for the primary (O3If*/WN6) and $70.4\pm9.3 M_\odot$ for the secondary (O3If*/WN5).  The system is almost in contact, and our preferred
solution has the primary filling its Roche surface, and the secondary almost filling its Roche surface. The secondary is  about 5,000~K hotter than the primary, and
we adopt effective temperatures of 37,000 and 42,000~K for the two
components.  Combined with the stellar radii determined from the light curve, this implies an absolute visual magnitude of $M_V=-7.3$,
consistent with the value we expect from our photometry and reddening estimate. 

\subsection{Is NGC~3603-A1 ``the most massive star known?"}

At one time, NGC~3603-A1 was touted as `the most massive binary."  While this is not completely correct, it may not be
completely wrong, either: we find that although there is a binary with more massive components known in the LMC, the mass of the primary in
NGC~3603-A1 is as large as any known star in the Milky Way that has a direct, Keplerian measurement.

  Although very high masses ($200-300M_\odot$) have been claimed for a variety of stars, such as R136a1 in the 30 Dor region, these measurements are based upon atmospheric and evolutionary models (see, e.g.,  \citealt{2010MNRAS.408..731C, 2020MNRAS.499.1918B}). While these studies are very important to our understanding of massive stars and the initial mass function, they
are not as fundamental as the measurement of masses directly using Kepler's laws in massive binaries. 

The LMC is known to contain a number of massive binaries, mostly in the 30 Dor cluster.  Mk 39 is the current record-holder, in terms of mass, with minimum masses of $105\pm11M_\odot$ and $80\pm11M_\odot$ \citep{2025MNRAS.539.1291P}. The orbital analysis required combining elements from an X-ray photometric solution along with optical data, resulting in a high eccentricity ($e=0.62$) and long period (649 d). An earlier
study by \citet{2008MNRAS.389..806S} of the star appears to be inconsistent with the new data, as noted by \citet{2025MNRAS.539.1291P}. Mk 34 has also been described as the ``most massive binary" \citep{2019MNRAS.484.2692T}; the minimum masses $m\sin^3{i}$ are $65\pm7M_\odot$ and $60\pm7M_\odot$.  They infer very high masses ($\sim140M_\odot$) based upon comparing the stars' luminosities with evolutionary tracks, as the inclination $i$ is unknown; i.e., the 100+$M_\odot$ masses are not from a direct measurement.  \citet{2022MNRAS.510.6133B} described Mk 33Na as a very massive colliding-wind binary system. Although its luminosity suggests the primary has a high (evolutionary) mass $>80M_\odot$,, the orbital inclination appears to be unfavorable (and has not been measured) and
$m_{\rm pri} \sin^3{i}$ is only $20.0\pm1.2M_\odot$.  
R144 has been analyzed by \citet{2021AandA...650A.147S}, with a primary and secondary masses of $74\pm4M_\odot$ and $69\pm4M_\odot$; the inclination is modeled from light curve modulation due to wind-wind eclipses and the derived masses are strongly at variance with evolutionary expectations given the stars' luminosities. R145 is a double-lined system analyzed by \citet{2017A&A...598A..85S}; the derived masses, while large, have rather huge uncertainties ($M_1=53^{+40}_{-20} M_\odot$, $M_2=54^{+40}_{-20} M_\odot$).  Its low inclination (39$\pm$6) results in a large correction to the minimum masses, and is derived from polarimetry. The next closest ``most massive binary" is R136-038, with a primary mass of $56.9\pm 0.6M_\odot$ \citep{2002ApJ...565..982M}. Like NGC~3603-A1, this is an unevolved system, but with a much smaller mass.  Thus, only Mk39,  R144, and R136-038 are likely useful for comparing inferred evolutionary masses with Keplerian determinations at LMC-like metallicities. 
{

The situation is similar in the Milky Way, where there are a plethora of stars whose high luminosities infer very large masses. The most famous example is probably $\eta$ Car, whose mass is often quoted as being at least 90~$M_\odot$, a result that traces back to the modeling work of  \citet{1992A&A...262..153H}.  Higher mass stars are listed in Table 1 of \citet{WalbornO2},  but often the luminosities of such stars have been over-estimated, either due to uncertain distances or their being undetected multiples. For instance, the luminosity of Pismis 24-1 suggested a mass of 200-300$M\odot$ \citep{WalbornO2}, but was later found to be a triple system by \citet{2007ApJ...660.1480M}.  Indeed, \citet{2010MNRAS.408..731C} cautioned that the high mass inferred from the luminosity of R136a1 might be affected by
undetected companions.  (See \citealt{2023A&A...679A..36S} for the latest on this fascinating star.)  
In terms of Galactic binaries with hard measurements, the largest masses were found in WR21a, 
a high eccentricity system with a period of 31.7 d
\citep{2008MNRAS.389.1447N,2016MNRAS.455.1275T,2022MNRAS.516.1149B}.  The discovery of eclipses in TESS data by \citet{2022MNRAS.516.1149B} allowed an inclination to be found, leading to a mass of $93.2^{+2.2}_{-1.9} M_\odot$ for the O2.5If*/WN6 primary, and $52.9^{+1.2}_{-1.1}$ for the O3V((f*))z secondary. The next most massive system is that of WR20a \citep{2004ApJ...611L..33B,2005AandA...432..985R}, with masses of $82.7\pm5.5M_\odot$ and $81.9\pm5.5M_\odot$.  Like NGC~3603-A1, the components are each O2-3If*/WN5-6 stars. Pretty much tied for the distinction is F2 in the Arches cluster \citep{2018AandA...617A..66L}, with masses of $82\pm12M_\odot$ and $60\pm8M_\odot$.  The system is comprised, somewhat surprisingly, of  a WN8-9 hydrogen-rich WR and an
O5-6 Ia$^+$.   

Thus, we conclude that NGC~3603-A1 is one of the most massive binaries known, with masses of $93.3\pm11.0M_\odot$ and $70.4\pm9.3M_\odot$ for the primary and secondary.    In the Milky Way, only WR21a is known to have an equally massive primary. In Table~\ref{tab:masses} we list the spectral types and masses of the highest mass binaries with well-determined parameters.

\begin{deluxetable}{l c c c c}
\tablecaption{\label{tab:masses}High Mass Binaries with Keplerian Mass Values}
\tablehead{
\colhead{Star}
&\colhead{Spectral Types}
&\colhead{M$_{\rm pri} (M_\odot$)}
&\colhead{M$_{\rm sec} (M_\odot$)}
&\colhead{Reference}
}
\startdata
\cutinhead{Milky Way} 
N3603-A1     & O3If$^*$/WN6 + O3If$^*$/WN5 & $93.3\pm11.0$ & $70.4\pm9.3$ & 1 \\
WR 21a         & O2.5If$^*$/WN6 +   O3V((f$^*$))z & $93.2^{+2.2}_{-1.9}$ & $52.9^{+1.2}_{-1.1}$ & 2 \\
WR 20a       & O3If$^*$/WN6 + O3If$^*$/WN6 & $82.7\pm5.5$  & $81.9\pm5.5$ & 3,4 \\
Arches-F2    & WN8-9h       + O5-6Ia$^+$   & $ 82\pm12$    & $60\pm8$     & 5 \\
\cutinhead{LMC}
Mk 39        & O2.5If*/WN6 + O3V-III       & $105\pm11$     & $80\pm11$    & 6 \\
R144         & WN5/6h      + WN6/7h        & $74\pm4$       & $79\pm4$     & 7 \\     
R136-038     & O3V         + O6V           & $56.9\pm0.6$   & $23.4\pm0.2$      & 8 \\
\enddata
\tablerefs{1--This paper; 2--\citealt{2022MNRAS.516.1149B}; 3--\citealt{2004ApJ...611L..33B}; 4--\citealt{2005AandA...432..985R}; 5--\citealt{2018AandA...617A..66L}; 
6--\citealt{2025MNRAS.539.1291P}; 7--\citealt{2021AandA...650A.147S}; 8--\citealt{2002ApJ...565..982M}.}
\end{deluxetable}
}

\subsection{Evolutionary Status}
\label{Sec-evol}

The NGC~3603-A1 system consists of a two very luminous, massive components almost in contact.  The more massive star is filling its Roche surface, with the secondary nearly filling its.   The more massive star (which we are calling the primary) has a (current) mass of $93.3\pm11.0 M_\odot$ and may have already evolved to a slightly cooler temperature, 37,000~K.  It is filling its Roche surface, and has been transferring mass to the secondary.  The secondary has a (current) mass of $70.4\pm9.3 M_\odot$ and an effective temperature of 
42,000~K.  Both of these stars have ``slash" WR spectral types, an O3If*/WN6 in he case of the primary, and a slightly higher excitation O3If*/WN5 in the case of the secondary.  Their spectra are indicative of an optically thick stellar wind.  

Our preference for the fit with the primary filling its Roche surface comes about for three reasons.  First, we expect that the primary will
be the first to fill its Roche surface, simply because the more massive star will evolve faster.  This is also consistent with the primary having
a slightly lower temperature as it begins its evolution.
Secondly, there can be little doubt that the
secondary has been spun up from accreting mass, and, along with it, angular momentum. It is clear from Figure~\ref{fig:example} that the rotational velocity of the secondary is much greater than that of the primary.  Correcting our measured $v\sin{i}$ values (280 km s$^{-1}$ and
500 km s$^{-1}$)
 for the system's inclination, we find rotational speeds
 of 290$\pm40$ km s$^{-1}$ for the primary, and 520$\pm$45 km s$^{-1}$ for the secondary.  (For the secondary, this is 37-39\% of its breakup speed.)  For either solution the primary is likely rotating synchronous with the orbit.  Using the radius of the primary derived from the solution in which it is filling its Roche surface, synchronous rotation would correspond to 303 km s$^{-1}$, in good agreement with our measurement.  Using the radius of the primary derived from the solution in which it is the secondary which is filling its Roche surface, synchronous rotation for the primary would correspond to 266 km s$^{-1}$, which is still within the uncertainty of our measurement.  However, the secondary is spinning much faster than its synchronous speed: for its radius derived from our preferred solution, its rotation speed would be 246 km s$^{-1}$, and for the other solution, 266 km s$^{-1}$.  Both of these values are much smaller than the measured 520$\pm$45 km s$^{-1}$ value.   So, regardless, the secondary is spinning much faster than is
 expected, considerably higher than all by the fastest O-type stars (see, e.g.,  \citealt{1996ApJ...463..737P,2013A&A...560A..29R}).  
 This strongly suggests that the secondary has been spun up, which argues that it is the primary that is filling its Roche surface and transferring
 mass (and angular momentum) to the secondary.
  
 The third reason has to do with the bolometric luminosities of the two stars. If we adopt a flux ratio of 0.794 (secondary to
primary) at 5570~\AA,   then the two components have $M_V=-6.7$ (primary) and $M_V=-6.4$ (secondary) for a combined absolute visual magnitude of $-7.3$.  
If we then adopt bolometric corrections of $-3.5$ and $-3.9$ for the
primary and secondary, respectively \citep{2005A&A...436.1049M}, we calculate very similar bolometric luminosities for the two components, each $\log L/L_\odot=6.0$. However, in the solution where the secondary fills its Roche surface and the
primary does not, the flux ratio at 5570~\AA\ would be 1.22.  That, combined with the higher temperature of the secondary, would make the
secondary much more luminous than the primary.  The absolute visual magnitudes would be $M_V=-6.3$ and $-6.6$ for the primary and
secondary, and the resulting bolometric luminosities would be $\log L/L_\odot=5.8$ and 6.5 for the primary and secondary, respectively.
  In both cases
the secondary would be over-luminous for its mass, but this is not totally unexpected given that accretion would force the star to expand and
drive it out of thermal equilibrium. (We are indebted to Dr.\ Jan Eldridge for offering this explanation.)  This may also explain why it is the hotter of the two stars, although as we note above, it may simply be that the primary has begun to evolve to slightly cooler temperatures. However, we expect the luminosity of the
primary to be  $\log L/L_\odot \sim 6.0$, and thus this solution would have the primary significantly under-luminous for its mass (see, e.g., \citealt{Sylvia}).
In either solution, wind-fed mass transfer to the secondary might also have played a role (see, e.g., \citealt{2021PASA...38...56H}).

One of the most interesting things, however, is that both the primary and secondary have optically thick winds with spectra dominated by broad WR-like emission.  The luminosities at which stars develop such winds and form O2-3If*/WN5-6 stars has never been
established observationally.  This is true for the other high-mass binaries, Mk 39, WR21a, and WR20a, as we see in Table~\ref{tab:masses}.  (We speculate as to whether the R144 components have been misclassified, and that these are also slash stars.) It is unclear whether this is purely due to their high luminosities, or if binary interactions are also contributing to the formation of optically thick winds.

NGC~3603-A1 has proven to not be an easy system to study; it has been over 40 years since its discovery as a spectroscopic binary by \citet{1984ApJ...284..631M}.   Significant progress was made by speckle imaging of \citet{1986AA...167L..15H} and the AO spectroscopy of \citet{2008MNRAS.389L..38S}, but it has really taken space-based
spectroscopy and photometry to reach our current understanding of this very massive binary.   With cuts to NASA space science looming, it serves as yet
one more example of the sort of study that may soon no longer be possible.

\begin{acknowledgements}

Lowell Observatory sits at the base of mountains sacred to tribes throughout the region. We honor their past, present, and future generations, who have lived here for millennia and will forever call this place home.  

This research is based on observations made with the NASA/ESA Hubble Space Telescope obtained from the Space Telescope Science Institute (STScI), which is operated by the Association of Universities for Research in Astronomy, Inc., under NASA contract NAS 5–26555. These observations are associated with programs GO-10602 (PI: Maiz Apellaniz), GO-11626 (PI: Massey), GO-12615 (PI: Schnurr), and GO-17527 (PI: Massey).   We thank our HST/STIS program scientist Alex Fullerton and our program coordinator Christian Soto for their help throughout the process of obtaining our data.  

Support for programs GO-11626 and  GO-17527 were provided to P.M. by NASA through a grant from the STScI. Support for S.B.'s participation via the National Science Foundation's Research Experiences for Undergraduates was made possible through NSF awards 1852478 and 1950901.  Additional support was also provided to P.M. through AST-2307594.

We thank Dr.\ Jan Eldridge for interesting correspondence on this system.  We also thank Mike Fitzpatrick for help with several IRAF-related issues.  NOIRLab IRAF is distributed by the Community Science and Data Center at NSF NOIRLab, which is managed by the Association of Universities for Research in Astronomy (AURA) under a cooperative agreement with the NSF.  Assistance in Python programming (both for error propagation and disentanglement) was provided by ChatGPT-4 \citep{2023arXiv230308774O}. An anonymous referee helped improve our presentation.

This work also made use of data from the European Space Agency (ESA) mission
{\it Gaia} (\url{https://www.cosmos.esa.int/gaia}), processed by the {\it Gaia}
Data Processing and Analysis Consortium (DPAC,
\url{https://www.cosmos.esa.int/web/gaia/dpac/consortium}). Funding for the DPAC
has been provided by national institutions, in particular the institutions
participating in the {\it Gaia} Multilateral Agreement.

\end{acknowledgements}

\facilities{HST (STIS, ACS), Gaia}
\bibliographystyle{aasjournalv7}
\bibliography{masterbib.bib}

\end{document}